\documentstyle[aps,prb,multicol]{revtex}
\begin{document}

\newcommand{\inc}{>}
\newcommand{\out}{<}
\draft

\title{Quantum interference in nanometric devices: \\
ballistic transport across arrays of T-shaped quantum wires}

\author{Guido Goldoni, Fausto Rossi, Elisa Molinari}

\address{Istituto Nazionale per la Fisica della Materia (INFM), and \\
Dipartimento di Fisica, Universit\`a di Modena, Via Campi
213/A, I-41100 Modena, Italy}

\date{\today}
\maketitle

\begin{abstract}
We propose that the recently realized T-shaped semiconductor quantum
wires (T-wires) could be exploited as three-terminal quantum
interference devices. T-wires are formed by intersecting
two quantum wells (QWs). By use of a scattering matrix
approach and the Landauer-B\"uttiker theory, we calculate the
conductance for ballistic transport in the parent
QWs and across the wire region
as a function of the injection energy.
We show that different conductance profiles
can be selected by tailoring the widths of the QWs and/or
combining more wires on the scale of the Fermi wavelength. Finally, we
discuss the possibility of obtaining spin-dependent conductance
of ballistic holes in the same structures.
\end{abstract}


\begin{multicols}{2}

\narrowtext

T-shaped quantum wires (T-wires) are semiconductor structures where
quasi-one-dimensional (q1D) confinement is achieved at the
intersection between two quantum wells (QWs).~\cite{Twires} T-wires
are obtained by first growing a GaAs/Al$_x$Ga$_{1-x}$As\ superlattice
(labelled QW1) on a (001) substrate; after cleavage, a GaAs QW
(labelled QW2) is grown over the exposed (110) surface, resulting in
an array of T-shaped regions where electron and hole wavefunctions can
be confined on a scale of few (5-10) $nm$. Up to now, the intensive
investigation of these structures focussed on optical properties,
and demonstrated strong one-dimensional quantum confinement of the lowest
excitonic transitions~\cite{Twires-opt} as well as evidence of laser
emission.~\cite{Twires-las} Transport experiments along the wires were
first obtained very recently.~\cite{Twires-transp}

At difference with wires obtained by other techniques, such as 
V-shaped or deep-etched wires,~\cite{review} 
the section of a T-wire
has an open geometry. Therefore, in addition to transport along the quantum 
wire in the q1D bound states, parallel transport in the constituent QWs
and across the wire region becomes possible if the two-dimensional (2D)
continuum is contacted (for example through the overgrown layer).
In addition to q1D bound states, falling below the 2D continuum edge
of the parent QWs, q1D resonant states exist within the 2D
continuum.~\cite{Goldoni97}
The injected carriers that travel ballistically
over the wire region ($nm$ scale) will show a strongly energy
dependent transmission, as a consequence of quantum interference
effects induced either by resonant q1D states or by the interplay
between the propagating modes of the parent QWs.

In semiconductors, quantum interference effects are normally achieved
in channels defined by gating an underlying 2D high mobility electron
gas with electrostatic potentials. Structures of this type
with T-shaped geometries have been proposed to achieve
device functions;~\cite{Sols89} in this case, the conductance along a
channel can be controlled by modulating the length of a lateral,
closed arm (stub).~\cite{stubs,nota-wires} In the present
T-wires, instead, the lateral arm (QW1) is open, and the conductance
is controlled by modulating the chemical potential (i.e. the injection
energy); as we will show, different shapes of the conductance as a function
of energy can be selected by tailoring the widths of the QWs and/or
combining more wires. 
In the proposed experiment with T-wires, the interference patterns
should be stable in a much larger temperature range than in previously
proposed structures, due to the $nm$-size confinement and the
large inter-subband splittings in the parent QWs (of the order of 0.1
eV); furthermore, the confinement is provided by high-quality 
interfaces, as demonstrated by the 
small excitonic linedwidths.~\cite{Twires-opt}

In the following we calculate the ballistic conductance for parallel
transport in the QWs through the T-wire intersection. 
Assuming perfect barrier confinement in the
parent QWs, the calculation of the
scattering matrix~\cite{Merz} presents no conceptual difficulty.
We divide the sample in four regions (see inset of Fig.~\ref{fig:1T});
in each region the wavefunction of energy $E$ is written as a linear 
combination of the
propagating and evanescent modes of the corresponding QW.
Indicating with $E_{1,n}$ and $f_{1,n}(x)$ the subband energies and
envelope functions
of QW1, and  with $E_{2,n}$ and $f_{2,n}(y)$ those of QW2, we have,
for zero in-wire momentum,
\begin{mathletters}
\begin{eqnarray}
\psi_A & = & \sum_n f_{2,n}(y)
         \left[a_n^\inc e^{i\xi_n x}+a_n^\out e^{-i\xi_n x}\right] ,
\label{psiA}
\\
\psi_B & = & \sum_n f_{1,n}(x)
         \left[b_n^\out e^{i\eta_n y}+b_n^\inc e^{-i\eta_n y}\right] ,
\label{psiB}
\\
\psi_C & = & \sum_n f_{2,n}(y)
         \left[c_n^\out e^{i\xi_n x}+c_n^\inc e^{-i\xi_n x}\right] ,
\label{psiC}
\\
\psi_D & = & \sum_n f_{2,n}(y)
         \left[d_n^+ e^{i\xi_n x}+d_n^- e^{-i\xi_n x}\right] + \nonumber\\
        & & ~~\sum_n f_{1,n}(x)
             e_n \left[e^{i\eta_n y}-e^{-i\eta_n y}\right] ,
\label{psiD}
\end{eqnarray}
\end{mathletters}
\noindent
where the wavevectors $\xi_n,\eta_n$ are given by
$\xi_n^2 = \left(\epsilon-n^2\right)\left(\pi/L_2\right)^2$ and
$\eta_n^2=\left(\epsilon-n^2/\alpha^2\right)\left(\pi/L_2\right)^2$.
Here,
$\epsilon=E/E_{2,1}$ is the energy in units of the lowest mode of QW2,
$E_{2,1}=\hbar^2\pi^2/2mL_2^2$, and $\alpha=L_1/L_2$. The two equations 
obtained at each interface  by matching both $\psi$ 
and its normal derivative are projected over the $n$th mode (i.e., multiplied
by the appropriate sine or cosine function and integrated over the
interface) and finally summed and subtracted to
obtain two new equations, relating either the
incoming or the outcoming wave coefficient through that interface to the
coefficients of the inside region D. Including $N$
modes~\cite{nota-eqs}
in the sums in (\ref{psiA})-(\ref{psiD}), and defining the vector
${\bf a}^\inc=(a^\inc_1,a^\inc_2,\ldots )$, and analogously  
for the other coefficients, we obtain a set of linear equations 
of the form 
\begin{mathletters}
\begin{eqnarray}
{\bf a}^{{}^\inc_\out} & = & {\bf d}^\pm + {\bf P}^\pm \cdot {\bf e}\,, 
\label{aio}\\
{\bf b}^{{}^\out_\inc} & = & {\bf V}^\pm \cdot {\bf d}^+ +
                   {\bf W}^\pm \cdot {\bf d}^- \label{co} + {\bf e}\,,
                   \label{bio}\\
{\bf c}^{{}^\out_\inc} & = & {\bf d}^\pm + {\bf Q}^\pm \cdot {\bf e}\,.
\label{cio}
\end{eqnarray}
\end{mathletters}
\noindent
where the eight $N\times N$ matrices
${\bf P}^\pm$, ${\bf Q}^\pm$, ${\bf V}^\pm$, and ${\bf W}^\pm$
ensue from the matching conditions and depend
on the geometrical parameters and on the energy.
By defining the incoming
and outcoming states $\left|in\right.\rangle=({\bf a}^\inc, {\bf
b}^\inc, {\bf c}^\inc)$ and
 $\left|out\right.\rangle=({\bf a}^\out, {\bf b}^\out, {\bf c}^\out)$,
and appropriate  $3N\times3N$ matrices ${\bf F,G}$ in terms of the
eight matrices above, 
the Eqs.~(\ref{aio})-(\ref{cio}) can be rewritten as 
\begin{equation}
|out\rangle={\bf F} \cdot \left({\bf e}, {\bf d}^+, {\bf d}^-\right)^T,
|in \rangle={\bf G} \cdot \left({\bf e}, {\bf d}^+, {\bf d}^-\right)^T.
\label{inout}
\end{equation}
Combining 
Eqs.~(\ref{inout}) we finally get
\begin{equation}
|out\rangle = {\bf F}\cdot{\bf G}^{-1} \cdot|in\rangle = {\bf
S}\cdot|in\rangle\,.
\label{eq:last}
\end{equation}
Equation (\ref{eq:last})
defines the scattering matrix ${\bf S}$, which gives at the same time
the bound state ($\epsilon<1$),~\cite{nota-eps}
satisfying the equation
$\det\left({\bf S}^{-1}\right)=0$, and the scattering states
($\epsilon>1$).

Since the scattering matrix is a property of the potential at a given
energy $\epsilon$, it allows to calculate all transmission coefficients,
say from mode $n$ in arm $A$ to mode $m$ in arm $C$,
by the same matrix, choosing the appropriate state $|in\rangle$;
to keep on with the example of $A\rightarrow C$
transmission, the transmission coefficient is
$t^{AC}_{n,m}=|c_n^</a_m^>|^2 \xi_n/\xi_m$.
In the following we shall concentrate on straight (i.e.,
$A\rightarrow C$) transmission along
QW2. We consider a configuration in which arm B is kept at the same
potential of arm C ($V_B=V_C$). Therefore, no carrier is injected into the structure
through arm B, and ${\bf b}^\inc=0$.~\cite{nota-stubs} Using one of the 
Eqs. (\ref{bio}), we can eliminate
the coefficients $\bf e$ from the equations and we can rewrite
(\ref{aio}),(\ref{cio}) as
\begin{equation}
\left(\begin{array}{c} {\bf c}^\out \\ {\bf c}^\inc \end{array}\right) =
{\bf T}\cdot
\left(\begin{array}{c} {\bf a}^\inc \\ {\bf a}^\out \end{array}\right),
\label{Tmatix}
\end{equation}
where $\bf T$ is the $2N\times2N$ $A\rightarrow C$
transfer matrix.
Note that if $V_B>V_C$, a case which we shall not investigate here,
a certain amount of charge would be inchoerently injected in
the system through arm B, and the transfer matrix would then contain an
inchoerent part which, in a three-teminal device, has been
discussed by B\"uttiker~\cite{Buttiker88}; in
Ref.~\onlinecite{Buttiker88} the ratio between the coherent and
incoherent parts of a two-terminal conductance
is modulated through the tunneling probability into a third,
randomizing terminal. The present T-shaped wires with $V_B>V_C$ 
might in fact be a system
to implement such an experiment, with the tunneling probability into arm B
being adjusted through the injection energy.

Going back to the $V_B=V_C$ case, the two-terminal 
Landauer-B\"uttiker (LB) conductance~\cite{LB} is
\begin{equation}
g = \frac{2e^2}{h} \sum_{i,j} t^{AC}_{i,j},
\end{equation}
In Fig.~\ref{fig:1T} we show the dimensionless conductance
$g/\frac{2e^2}{h}$ as a function of the energy $\epsilon$ and for
selected values of the parameter $\alpha$. We recognize two types
of behaviours: for samples in which the width of side arm, QW1, matches
an integer number of semiperiods of the incoming wave
($\alpha=0.5$, $1$, $1.5$, left panels) there are
strong reflection resonances~\cite{Shao94} at the energies of resonant q
1D states
localized at the intersection; when these states appear, their energy
is at or sligtly below the onset of a new propagating state along QW2
which, in the present units, is at $n^2=1,4,9,\ldots $.
When the matching condition is not
fulfilled ($\alpha=0.75$, $1.25$, $1.75$, right panels), the
conductance shows, on top of
a regular increase, a square-wave behaviour, with sudden drops and
rises when new propagating channels open in QW1 or QW2, i.e, the
current coming from arm A flows into the side arm B or into the
straight arm C,
depending on the energy.

If successive wires in an array are at a distance larger than the
coherence length, incoherent scattering will redistribute carriers
homogeneously among
the propagating modes; therefore, the conductance of $N$ incoherently
coupled wires is $G^N$, apart
from possible broadening due to fluctuations
in the QW widths
on the monolayer scale; this would not wash out completely the interference
patterns, however, as long as the intersubband splittings are large.
Conversely, the LB conductance of a single wire
can be changed by coupling more wires on the scale of the
Fermi wavelength; this possibility is a distinct advantage of
structures grown by epitaxy. As an example, we consider
two coupled T-wires with a barrier of width $L_d$
between two QW1s (see inset in Fig.~\ref{fig:2T}). The
transmission coefficient of the whole structure can be easily calculated,
as the total $\bf T$-matrix is the
product of the $\bf T$-matrices of the isolated wires. In
Fig.~\ref{fig:2T}(a) we compare, for the case $\alpha=0.5$, the
conductance of a single wire with the conductance of two coupled wires
for $L_b=L_1$ and $L_b=2L_1$. In the first case, the conductance shows
a double resonance, which is a fingerprint of the bonding and
antibonding combinations of the resonant q1D state of the isolated
wires. In the second case, instead, the resonance is completely
suppressed.
In Fig.~\ref{fig:2T}(b) we compare, for the
case $\alpha=1.2$, the single and coupled wire conductance with
$L_b=L_1/2$ and $L_b=2 L_1/3$. The coupled wire case shows sharper
modulations with respect to the isolated wire case.

Finally, we consider the possibility of transmitting holes, instead of
conduction electrons, through a T-wire. The valence subbands of a
T-wire are strongly spin-split at finite in-wire
wavevectors.~\cite{Goldoni97} This can be understood in the following
way: if the parent QWs of a T-wire were isolated, the spin-degenerate
valence subbands would be degenerate at some finite in-wire
wavevector, because of the different effective masses for (001)- and (110)-
grown wells. In each QW valence states can be characterized
with the component of the total angular momentum $J=3/2$, the $J$
quantization axis being along the growth direction.~\cite{Bastard}
Therefore, a state with a well defined component $J_z$, say $J_z=3/2$
in one QW, is a mixture of $J_z=\pm 3/2,\pm 1/2$ states in the other
QW; as a consequence, the strong spin-orbit coupling of valence
states, by coupling heavy hole (i.e., $J_z=\pm 3/2$) and light hole
(i.e., $J_z=\pm 1/2$) states removes the degeneracy and results in a
large spin-splitting.~\cite{Goldoni97} 
Therefore, if holes cross the wire region having
a finite component of the wavevector along the wire axis, the
transmitted current at selected energies could be strongly
spin-polarized.

In summary, we have proposed a nanostructure interference device
based on cleaved-edge-overgrown T-shaped quantum wires, and shown that
its conductance profile can be tailored by chosing appropriate widths
of the constituent QWs and of the barriers between adjacent
structures. Possible applications as spin-selective devices for holes
were also proposed, and will require further theoretical and experimental 
investigations.

\begin{figure}
\caption{Two-terminal conductance vs injection energy for selected values
of $\alpha$, according to the labels. The relevant geometric parameters are
defined in the inset.}
\label{fig:1T}
\end{figure}

\begin{figure}
\caption{Two-terminal conductance vs injection energy for a
single wire and for two coupled wires for
a) $\alpha=0.5$ and  b) $\alpha=1.2$ and for selected values of $L_b$.
The relevant geometric parameters are defined in the inset.}
\label{fig:2T}
\end{figure}


\end{multicols}

\begin{references}
\bibitem{nota-wires} Note the different dimensionality of the devices
obtained by gating a 2D electron gas with respect to the 
device proposed here: 
In Refs.~\protect\onlinecite{Sols89,stubs} the lateral arms of the T
are q1D channels, therefore termed ``wires''. In 
T-wires obtained by cleaved edge overgrowth, 
the lateral arms of the T are sections of 2D systems, 
and the term ``wire'' refers to the q1D nature of the states 
localized at the intersection of such QWs.
\bibitem{nota-eqs} In principle, Eqs. (\ref{aio})-(\ref{cio}) can only
be satisfied by an infinite number of modes, which means that we have to
check the convergence of our results with respect to $N$. In practice,
since the energy of each mode increases as $n^2$ in a infinite well,
very few modes are necessary.
\bibitem{nota-eps} For a wire with $L_1=L_2$, the bound state is always at 
$\epsilon=0.807$; when the parent QWs are different, $0.807<\epsilon<1$.
\bibitem{nota-stubs} If arm $B$ were terminated by a
perfectly reflecting wall at a distance $L$, as in
Refs.~\onlinecite{Sols89,stubs},
we should set $b^\inc_n=-b^\out_n \exp(2i\eta_nL)$ to account for the 
phase accumulated going along arm B.
\end{references}
\end{document}